\documentclass{elsart}

\usepackage{graphicx}

\begin{document}
\begin{frontmatter}

\title{Phonon scattering in quasicrystalline \emph{i}-Al$_{72}$Pd$_{19.5}$Mn$_{8.5}$: A study of the low-temperature thermal conductivity}

\author[ST,ZG]{A. Bilu\v{s}i\'{c}\corauthref{corauth}}
\ead{bilusic@pmfst.hr}
\corauth[corauth]{Corresponding author. Tel: + 385 21 385 133; fax: + 385 21 385 431.}
\author[ZG]{A. Smontara}
\author[LJ]{J. Dolin\v{s}ek}
\author[LJ]{P. McGuiness}
\author[ZH]{H. R. Ott}

\address[ST]{Faculty of Natural Sciences, University of Split, N. Tesle 12, HR-21000 Split, Croatia}
\address[ZG]{Institute of Physics, P.O.B. 304, HR-10001 Zagreb, Croatia}
\address[LJ]{Jo\v{z}ef Stefan Institute, Jamova 39, SI-1000 Ljubljana, Slovenia}
\address[ZH]{Laboratorium f\"{u}r Festk\"{o}rperphysik, ETH H\"{o}nggerberg, CH-8093, Z\"{u}rich, Switzerland}

\begin{abstract}
We measured the thermal conductivity of an icosahedral quasicrystal \emph{i}-Al$_{72}$Pd$_{19.5}$Mn$_{8.5}$ in the temperature range between 0.4 K and 300 K. The analysis of the low-temperature results was based on a Debye-type model. The results  of the analysis for the two temperature regions of $0.4$ K $<T<40$ K and $0.4$ K $<T<1$ K, are not consistent in the sense that a tunnelling-states contribution to phonon scattering is verified only for $0.4$ K $<T<1$ K. The same fitting procedure indicates that structural defects of the stacking-fault type are an important source of phonon scattering. Their physical presence was cleary identified by a transmission electron microscopy experiment.
\end{abstract}

\end{frontmatter}

\section{Introduction}\label{sec:intro}

Quasicrystals are materials which exhibit long-range structural order without periodicity. This feature allows for the existence of 'forbidden' crystal symmetries. For example, icosahedral quasicrystals exhibit diffraction patterns with fivefold symmetry axes and for decagonal quasicrystals tenfold symmetry axes are observed. This particular character of the structure raises the question of how the dynamics of quasilattices is best described and whether the aperiodicity implies that the vibration spectra of quasicrystals are similar to those of amorphous materials. A particular characteristics of amorphous materials is the existence of so-called tunnelling states, introduced by Philips \cite{philips72} and Anderson \etal{} \cite{anderson72}. The tunneling-states model assumes the existence of structural configurations of the atoms which are almost degenerate in energy. These configurations or states are separated by low potential barriers through which "tunnelling" between these states is possible. The energy spread of these states is of the order of 1 K$\cdot k_{B}$ and hence their influence on physical properties is most significant at temperatures in the sub-Kelvin range. 

Many experimental attempts have been made in order to find evidence for tunnelling states in quasicrystals. Suitable experiments for this purpose are low-temperature measurements of the thermal conductivity and ultrasound propagation. The thermal conductivity data of Chernikov and co-workers \cite{chernikov95,chernikov98} were interpreted as indicating that below 1 K, the main phonon scatterers are tunnelling states. Experiments probing the sound-velocity variation in decagonal \emph{d}-AlNiCo quasicrystals (decagonal quasicrystals are built from quasicrystal line planes which are stacked periodically), indicated an anisotropy in the tunnelling-state density \cite{bert02} in the sense that it is four times larger in the quasiperiodic direction than it is in the direction along the periodic stacking. The same experiment made on \emph{i}-AlLiCu icosahedral quasicrystals \cite{bert01} showed that the importance of tunnelling-states is more pronounced in samples with a higher degree of quasicrystalline order. From these observations it was suggested that tunnelling states are intrinsic to quasicrystalline structures. Bert \etal{} measured ultrasound attenuation in icosahedral \emph{i}-AlCuFe \cite{bert00} and derived a density of  tunnelling-states  which is comparable with those of metallic glasses. Results of low-temperature thermal-conductivity and ultrasound measurements on a quasicrystal of the \emph{i}-AlPdMn family \cite{thompson00} demonstrated that the quasilattice dynamics is very similar to that of amorphous solids. Legault \etal{} \cite{legault98}, on the basis of thermal conductivity below 1 K, found that, besides tunnelling states, stacking-fault like defects have to be taken into account to explain the data. Ultrasound experiments on a single-grain quasicrystal of \emph{i}-MgZnY \cite{sterzel00} revealed features that were claimed to reflect glassy behavior. Thermal conductivity data measured on another single-grain sample of the same family, however, gave no evidence for the presence of tunnelling states \cite{gianno00}.

In order to further broaden the existing data base, we made measurements of the thermal conductivity of a member of the \emph{i}-AlPdMn family, covering a broad range of temperatures between 0.4 K and 300 K. The data were analyzed employing a Debye-type formula for the thermal conductivity. Results of the analysis were supported by the transmission electron microscopy experiment.

\section{Experimental}\label{sec:exp}
 
The sample with the composition \emph{i}-Al$_{72}$Pd$_{19.5}$Mn$_{8.5}$ was grown with a self-flux technique \cite{fisher99} at Ames Laboratory by I.R. Fisher and P.C. Canfield. Measurements of the thermal conductivity between 0.4 K and 300 K were performed at ETH Z\"{u}rich employing an absolute steady-state heat-flow technique. The sample was in the form of a bar, with dimensions 0.5$\times$0.6$\times$5.9 mm$^{3}$ and a mass of 7.3 mg. The thermal flux through the sample was generated by a 1 k$\Omega$ RuO$_{2}$ chip-resistor, glued to one end of the sample, while the other end was thermally anchored to a copper heat sink. Two gold wires, 50 $\mu$m in diameter, were attached to the sample and were used as thermal contacts for the thermocouple and the resistor thermometers (see below); EpoTek H20E was used as a glue. Between 5 K and 300 K, the thermal conductivity was measured in a helium-4 cryostat, using a 25 $\mu$m chromel-gold (doped with 0.07\% Fe) differential thermocouple. Between 0.4 K and 4 K, a helium-3 refrigerator was used as the cooling device. For monitoring the temperature gradient of the sample in this temperature regime, two 10 k$\Omega$ RuO$_{2}$, \emph{in-situ} calibrated resistors were used. As shown in Ref. \cite{gianno00}, heat losses due to radiation and conduction through the contact wires can be neglected with this set-up at temperatures below 300 K.  

Transmission electron microscopy experiments (TEM) were performed at the Jo\v{z}ef Stefan Institute, Ljubljana, Slovenia. For these purpose, material obtained from the same sample that was used for the thermal conductivity experiment was cut into 3-$\mu$m discs, mechanically ground to a thickness of 100 $\mu$m, and dimpled down to 20 $\mu$m in the disc centre. Transmissive regions in the specimen's centre were finally achieved by ion milling using 4 kV Ar$^{+}$ at an incidence angle of 12$^{\mathrm{o}}$. Structural features of the sample were examined in a Jeol, 2000FX microscope, which was operated at 100 keV and a beam current of about 0.25 nA. During specimen observation no damage of the sample by the electron beam and no evidence for significant specimen contamination were noticed.

\section{Results and Discussion}\label{sec:results}
\subsection{Electronic thermal conductivity}
The measured temperature dependence of the thermal conductivity $\kappa(T)$ of \emph{i}-Al$_{72}$Pd$_{19.5}$Mn$_{8.5}$ is shown as open circles in the main panel of Figure 1. The value of $\kappa(T)$ at 300 K is 4.6 W/mK. Due to the uncertainty in controlling the sample geometry, the uncertainty of this value is of the order of $\pm$10\%. This rather low value of the thermal conductivity is typical for quasicrystals, and is a consequence of the non-periodicity of the lattice, which hampers the itineracy of lattice excitations, except for those with small wave vectors. The electronic contribution to the thermal conductivity $\kappa_{el} (T)$ is usually estimated by employing the law of Wiedemann and Franz (WF) \cite{ashcroft-mermin} in the form

\begin{equation}\label{eq:WFL}
    \kappa_{el}^{WF}(T)=\frac{\pi^{2}}{3} \frac{k_{B}^{2}}{e^{2}} \sigma(T) T \mathrm{,}
\end{equation}

\noindent
where the prefactor $L_{0} = \big( \pi^{2}/3 \big) \big( k_{B}^{2}/e^{2} \big)$ is the Lorenz number, $\sigma(T)$ is the electrical conductivity and $T$ is the temperature. The above formula is valid for elastic scattering. \emph{Ab initio} calculations of the Lorenz number for \emph{i}-AlCuFe quasicrystals \cite{landauro01} indicated that at room temperature it is about 40\% larger than $L_{0}$ and, with decreasing temperature, approaches the value of $L_{0}$. To calculate $\kappa_{el}^{WF} (T)$, we used the electrical conductivity data obtained from corresponding measurements on the same sample \cite{dolinsek02}. A comparison with the measured $\kappa(T)$ data showed that the electronic contribution is only of the order of 1\% and therefore any deviations from the WF behaviour are insignificant. 

\subsection{Quasilattice thermal conductivity}\label{subsec:Debye}
The quasilattice contribution to the thermal conductivity $\kappa_{qlatt}(T)$ is obtained by subtracting the calculated electronic part from the measured total thermal conductivity and hence

\begin{equation}\label{eq:Kqlatt}
    \kappa_{qlatt}(T)=\kappa(T)-\kappa_{el}(T) \mathrm{.}
\end{equation}

Since the quasicrystalline structure is not periodic, only long-wavelength pho\-nons can be defined in quasilattices with some confidence. Inelastic neutron (INS) \cite{janot93,boissieu93,boudard95} and X-ray (IXS) \cite{krisch02} scattering experiments on samples of the \emph{i}-AlPdMn quasicrystal family revealed the existence of isotropic acoustic phonon modes with wave vectors $q<0.3$ \AA{} and energies $E<6$ meV. For energies higher than 12 meV, dispersionless vibrational states were identified. The energy of 6 meV corresponds to a thermal energy of approximately 50 K$\cdot k_{B}$ and thus justifies the use of the Debye-type approximation for analyzing the thermal conductivity data at temperatures less than 50 K. The relevant Debye-type equation reads \cite{berman}

\begin{equation}\label{eq:Debye}
    \kappa_{Debye}=\frac{k_{B}}{2 \pi^{2} \overline{v}} \bigg( \frac{k_{B}}{\hbar} \bigg) ^{3} T^{3} \int_{0}^{\theta/T} \tau (x) \frac{x^{4} e^{x}}{\big( e^{x} -1 \big) ^{2}}dx \mathrm{,}
\end{equation} 

\noindent
where $k_{B}$ and $\hbar$ are Boltzmann's and the reduced Planck constant, respectively; $\overline{v}$ is the average sound velocity, $\theta$ the Debye temperature, $\tau(x)$ the phonon relaxation time and $x = \hbar \omega / k_{B} T$, where $\hbar \omega$ is the phonon energy. Ultrasonic data \cite{amazit92} provided values for the transversal ($v_{T}=3593$) m/s) and the longitudinal ($v_{L}=6520$) m/s) sound velocities. Since neutron and X-ray scattering experiments have shown that the excitation spectra of vibrational states in this alloy are isotropic, the average sound velocity $\overline{v}$ can be obtained using the relation

\begin{equation}\label{eq:velocity}
    \frac{3}{\overline{v}^{3}}=\frac{1}{v_{L}^{3}}+\frac{2}{v_{T}^{3}}\mathrm{.}
\end{equation} 

\noindent
From this equation $\overline{v}=4004$ m/s. The Debye temperature $\theta$, was calculated by Li and Liu \cite{li01}, based on the specific heat data of W\"{a}lti \etal{} \cite{waelti98} . They obtained $\theta = 492$ K.

In equation (\ref{eq:Debye}) the different phonon scattering processes are incorporated in the relaxation time $\tau(x)$.  As usual, we assume that Matthiessen's rule is valid, and the mutual independence of the scattering processes is reflected in

\begin{equation}\label{eq:Matthiessen}
    \tau^{-1}=\sum_{i} \tau^{-1}_i \mathrm{,}
\end{equation} 

\noindent
where $\tau^{-1}_{i}$ is a scattering rate related to the $i$-th scattering channel. At the lowest temperatures reached in our experiment, the Casimir scattering, i.e., scattering of phonons at sample boundaries may contribute significantly to the total phonon relaxation rate. The corresponding scattering rate is given by the phonon mean free path $l$ and the average sound velocity $\overline{v}$ by

\begin{equation}\label{eq:Casimir}
    \tau_{Cas}^{-1}=\frac{\overline{v}}{l} \mathrm{.}
\end{equation} 

\noindent
Inserting (\ref{eq:Casimir}) into equation (\ref{eq:Debye}) leads to $\kappa_{Debye}(T) \propto T^{3}$, if this scattering process is dominating.

At the temperatures defined by the energy spread of tunneling states ($ \approx 1$ K$\cdot k_B$), transitions between states that are separated by a shallow barrier may be induced by phonons. This type of phonon scattering is absent in crystals and this is the reason why at low temperatures, the temperature dependences of the lattice thermal conductivities of crystals and amorphous solids differ in a very characteristic way. Tunnelling states act as scattering centres for phonons, and the corresponding phonon scattering rate is given by \cite{philips72}

\begin{equation}\label{eq:ts}
    \tau_{ts}^{-1}=\frac{\overline{P} \gamma ^{2}}{\rho \overline{v}^{2}} \pi \omega \tanh \bigg( \frac{\hbar \omega}{2 k_{B} T} \bigg) \mathrm{,}
\end{equation} 

\noindent
where $\overline{P}$ is the density of tunnelling states, $\gamma$ the average phonon-tunnelling states coupling constant and $\rho$ the mass density. A quantity $C$, defined by

\begin{equation}\label{eq:tun_str}
    C \equiv \frac{\overline{P} \gamma ^{2}}{\rho \overline{v}^{2}} \mathrm{,}
\end{equation} 

\noindent
is denoted as the tunnelling strength. Inserting equation (\ref{eq:ts}) into the integral in equation (\ref{eq:Debye}), leads to $\kappa_{Debye} (T)$ being proportional to $T^{2}$ for a dominating scattering of this type.

In order to obtain an optimal fit to the low-temperature thermal conductivity data of \emph{i}-AlPdMn, the authors of Ref. \cite{legault98} had to assume that the total scattering rate is the sum of two contributions, namely one according to equation (\ref{eq:ts}), and a second contribution that leads to a linear $T$ dependence of $\kappa_{Debye} (T)$. The latter term may be caused by the scattering of phonons at stacking faults, with a scattering rate that is given by \cite{klemens}

\begin{equation}\label{eq:sf}
    \tau_{sf}^{-1}=\frac{7}{10} \frac{a^{2}}{\overline{v}} \Gamma ^{2} \omega ^{2} N_{s} \mathrm{,}
\end{equation} 

\noindent
where $a$ is a lattice constant, $\Gamma$ the Gr\"{u}neisen parameter and $N_{s}$ the linear density of stacking faults. We thus assume that the scattering process that is represented by equation (\ref{eq:sf}) also needs to be considered in the analysis of our data.

Theoretical calculations of the vibrational excitation spectrum of a one-di\-men\-si\-o\-nal quasicrystal, i.e., a Fibbonacci chain \cite{kalugin96}, revealed a dense distribution of energy gaps. Although not proven rigorously, the same characteristic feature is expected for three-dimensional quasilattices. Together with the lack of well defined reciprocal lattice in quasicrystals, this leads to a power-law dependence of the generalized umklapp (\emph{quasiumklapp}) scattering rate \cite{kalugin96}

\begin{equation}\label{eq:qu}
    \tau_{qu}^{-1} \propto \omega ^{2} T^{4} \mathrm{.}
\end{equation} 

\noindent
In order to establish the dominant scattering processes, we attempted to fit the data of $\kappa_{qlatt}(T)$ using equation (\ref{eq:Debye}), considering all physically reasonable combinations of scattering rates, given by equations (\ref{eq:Casimir}), (\ref{eq:ts}), (\ref{eq:sf}) and (\ref{eq:qu}). According to the arguments concerning the applicability of equation (\ref{eq:Debye}) below 50 K, the fitting procedure was applied for temperatures below 40 K. The best fit was obtained with the assumption that the total relaxation rate $\tau^{-1}$ is given by the sum of three of the above cited scattering rates

\begin{equation}\label{eq:tau fit}
    \tau^{-1}=\tau^{-1}_{Cas} + \tau^{-1}_{sf} + \tau^{*-1}_{qu} \mathrm{,}
\end{equation}

\noindent
ignoring tunnelling states as relevant scattering centres. The term $\tau_{qu}^{*-1}$ represents a modified quasiumklapp scattering rate. The use of the quasiumklapp scattering rate (\ref{eq:qu}) proposed by Kalugin \etal{} \cite{kalugin96} did not give a satisfactory fit-curve in our case. This motivated us to test several power-law forms for the quasiumklapp scattering rate ($\propto \omega^2 T^2$, $\omega T^3$, $\omega^3 T$). The form $\tau^{*-1}_{qu} \propto \omega^2 T^2$ gives the best matching bet\-ween the fit and the measured curves of $\kappa_{qlatt}(T)$. Other forms of $\tau_{qu}^{*-1}$ were chosen by other authors, including $\tau_{qu}^{*-1} \propto \omega^{3} T$, which was used in reference \cite{gianno00} for the analysis of the thermal conductivity of \emph{i}-ZnMgY. The scattering rates of $\tau_{qu}^{*-1} \propto \omega^2 T^2$ and $\tau_{qu}^{*-1} \propto \omega^3 T$ lead to a $T^{-1}$-temperature dependence of the quasilattice thermal conductivity, while the scattering rate given by equation (\ref{eq:qu}) gives $\kappa_{qlatt}(T) \sim T^{-3}$. Deviations from the $T^{-3}$-law are also observed in quasicrystalline approximant phases \cite{dolinsek05,bihar06}.  Several reasons might cause the deviation from the $T^{-3}$-temperature dependence of the quasilattice thermal conductivity. First, at temperatures of few a tens of Kelvins there is the possibility of  hopping excitations of  localized quasilattice vibrations, whose activation energies is of the order of 1 meV. Second, IXS experiments \cite{krisch02} revealed the existence of low-lying vibrational states in \emph{i}-AlPdMn phases. Their excitation also may mask the temperature dependence of  $\kappa_{qlatt}$ that is consistent with $\tau_{qu}^{-1}$ as given by equation (\ref{eq:qu}).

Fitting equation (\ref{eq:Debye}) to the data of $\kappa_{qlatt}(T)$ allows for an evaluation of the mean free path $l$, and the stacking-fault density $N_{S}$. According to equation (\ref{eq:Casimir}) and $\overline v = 4004$ m/s, the fitting procedure gave the value of 0.8 mm for the phonon mean free path $l$ at low temperatures. To calculate $N_{S}$, the values of the 'lattice' constant $a$ and the Gr\"{u}neisen parameter $\Gamma$ are needed. In non-periodic structures, such as quasicrystals, the definition of a 'lattice' constant needs some explanation. The structure model of quasicrystals assumes that their structures are a low-dimensional projection of a structure that is periodic in a higher dimensional space. For example, icosahedral quasicrystals are obtained by the projection of a 6-D cubic lattice onto a 3-D space. Elser \cite{elser96} defined the 'lattice' constant $a$ of icosahedral quasicrystals as the extension of the projection of the six-dimensional cube edge on the 3-D space, multiplied by the golden mean $\tau$ ($\tau = (\sqrt{5}+1)/2$). The value of $a$ for \emph{i}-AlPdMn calculated in this way is 7.4 \AA{}. The Gr\"{u}neisen parameter can also be calculated from the data of $\Gamma$'s of the constituent elements \cite{gschneider64}, with the atomic composition fixing the weight factors. The relevant Gr\"{u}neisen parameter calculated in this way is $\Gamma = 2.1$. The experimental determination of the Gr\"{u}neisen parameter of the \emph{i}-AlPdMn family showed that above 50 K it adopts a constant value of around 1.8 \cite{kajiyima00,swenson04}. Low-temperature data of Swenson and co-workers \cite{swenson04} suggest an increase of the $\Gamma$-value up to 8 at temperatures of the order of 1 K. Because of this uncertainty only an order of magnitude estimate of $N_{S}$ can be made. The stacking fault density $N_{S}$ lies between 0.1 and 1 $\mu$m$^{-1}$. 

The sum $\kappa_{Debye}(T)+\kappa_{el}(T)$, where $\kappa_{Debye}(T)$ is the fit of $\kappa_{qlatt}(T)$ to equation (\ref{eq:Debye}) obtained for $0.4$ K$<T<40$ K and $\kappa_{el}(T)=\kappa_{el}^{WF}(T)$, is shown as a solid line on Figure 1 It may be seen that the calculated curve reproduces the measured data very well below $50$ K. Above 50 K the experimental data deviate from the sum $\kappa_{Debye}(T) + \kappa_{el}(T)$. The thermal activation of localized vibrational modes opens a channel for the heat transfer by hopping excitations, which generally can be described by an activation law

\begin{equation}\label{eq:activation}
    \kappa_{hop} (T) = \kappa_{hop}^{0} \cdot \exp (-E_{a} / k_{B}T) \mathrm{,}
\end{equation}
 
\noindent
where $E_{a}$ is the mean activation energy, and $\kappa_{hop}^{0}$ is a hopping-type thermal conductivity of the system at infinite temperature. The difference between the experimental data $\kappa(T)$ and the sum $\kappa_{Debye}(T) + \kappa_{el}(T)$ was attributed to the hopping-type heat-conduction channel and was fitted to Eq. (\ref{eq:activation}). The fit gave  $E_{a}=29$ meV and $\kappa_{hop}^{0} = 8.0$ W/mK. The curve of the total thermal conductivity

\begin{equation}\label{eq:kappa_tot}
    \kappa_{tot}(T) = \kappa_{Debye} (T) + \kappa_{el} (T) + \kappa_{hop} (T) \mathrm{,}
\end{equation}
 
\noindent
and the individual curves of $\kappa_{Debye}(T)$, $\kappa_{el}(T)=\kappa_{el}^{WF}(T)$, and $\kappa_{hop} (T)$ are presented in the Figure 2 as the solid, solid grey, dashed, and dotted-dashed lines, respectively. It is obvious that $\kappa_{tot}(T)$ reproduces the measured thermal conductivity $\kappa(T)$ remarkably well at all covered temperatures.

The fit of $\kappa_{qlatt}(T)$ to $\kappa_{Debye}(T)$ (equation (\ref{eq:Debye})) for $0.4$ K $<T<40$ K leads to the conclusion that tunnelling states are not a relevant source of phonon scattering. The same was concluded for \emph{i}-MgZnY in Ref. \cite{gianno00}, where the same type of the thermal-conductivity analysis was used below 50 K. Other authors, who observed that tunnelling states do take part in the phonon scattering \cite{chernikov95,chernikov98,thompson00,legault98}, made their Debye-type analysis of $\kappa(T)$ below 1 K. Existing experimental data besides thermal conductivity, i.e. results of acoustic experiments, show that tunnelling-state defects are intrinsic constituents of the quasicrystal structure. Thus, we face a discrepancy between fits of quasicrystal thermal conductivity data for $0.4$ K $<T<40$ K and other claims in the existing literature. In view of this situation, we also attempted to fit our quasilattice thermal conductivity data $\kappa_{qlatt}(T)$ to equation (\ref{eq:Debye}) in the temperature interval between 0.4 K (the lowest temperature reached in the experiment) and 1 K. The best fit is obtained by the following choice of phonon scattering rates

\begin{equation}\label{eq:tau fit lowT}
    \tau^{-1}=\tau^{-1}_{Cas} + \tau^{-1}_{sf} + \tau^{-1}_{ts} \mathrm{.}
\end{equation}

\noindent
The fit is presented as a solid line in the inset of Figure 1. The fit parameters are $l=1$ mm, and the order of magnitude of $N_{S}$ ranges between 1 and 10 $\mu$m$^{-1}$. Note that the uncertainty is due to the approximate evaluation of the Gr\"{u}neisen parameter $\Gamma$. The parameter $C$ is $0.2 \times 10^{-4}$. Existing experimental data on the tunnelling strength for the \emph{i}-AlPdMn quasicrystal family were collected not only by using thermal conductivity data, but also results of acoustic-attenuation and sound-velocity experiments were invoked \cite{thompson00,pohl02}, and the resulting values of $C$ range between $0.3 \times 10^{-4}$ and $3 \times 10^{-4}$.

We note two kinds of discrepancies between fits of $\kappa_{qlatt}(T)$ to $\kappa_{Debye}(T)$ for the temperature ranges $0.4$ K $<T<40$ K and $0.4$ K $<T<1$ K. The first one is related to the importance of tunnelling-states scattering which seems to depend on the chosen temperature range of data fitting. It might also be the case that the Debye-type model of the thermal conductivity is really valid only at very low temperatures. The second discrepancy between $0.4$ K $<T<40$ K-fit and $0.4$ K $<T<1$ K-fit concerns the value of $N_{S}$, which ranges between 0.1 and 1 $\mu$m$^{-1}$ for the first fit range and 1 and 10 $\mu$m$^{-1}$ for the second. A clarification is offered by the results of the TEM experiment, which provides information on the presence and the density of stacking faults in the sample. A TEM photograph of the investigated sample is shown in Figure 3, where some of the stacking-fault contrasts are indicated by white arrows. The order of magnitude of the linear stacking-fault density $N_{S}$ can be estimated from Figure 3 and lies between 1 and 10 $\mu$m$^{-1}$. This supports the results of the thermal conductivity analysis obtained for the temperature range $0.4$ K $<T<1$ K, indicating that the Debye-type analysis of $\kappa(T)$ for quasicrystals is more appropriate at very low temperatures.

\section{Conclusion}
In this work, we investigated the thermal conductivity of an icosahedral quasicrystal \emph{i}-Al$_{72}$Pd$_{19.5}$Mn$_{8.5}$ in the temperature range between 0.4 K and 300 K. The objective was to study the influence of tunnelling states on the phonon scattering affecting the low-temperature thermal conductivity, because some thermal conductivity data available in the literature do not corroborate the general claim of tunnelling states as a characteristic of a quasicrystal structure.

Fits to our thermal conductivity results were made using a Debye-type analysis in two temperature regions: $0.4$ K $<T<40$ K and $0.4$ K $<T<1$ K. The resulting fits are sensitive to the chosen temperature regime: for $0.4$ K $<T<40$ K the best fit suggests that tunnelling states are not significantly affecting the scattering of phonons, contrary to the majority of similar work available in the literature. The fit to the data in region between 0.4 K and 1 K, however indicates the relevance of tunneling states as phonon scattering centres. Besides the question of the existence of tunnelling states, the same fit confirms the importance of scattering centers in the form of stacking faults, which are actually observed in transmission electron microscopy experiments.  

\ack
A.B. acknowledges financial support from the Swiss Federal Commission for Foreign Students. The authors thank I.R. Fisher and P.C. Canfield for supplying the sample, and E. Tuti\v{s} for fruitful discussions. The work was in part finacially supported by the Swiss National Science Foundation and the Croatian Ministry of Science, Education and Sport.

\newpage
\begin{center}
\begin{figure}[t]
\includegraphics[width=120mm]{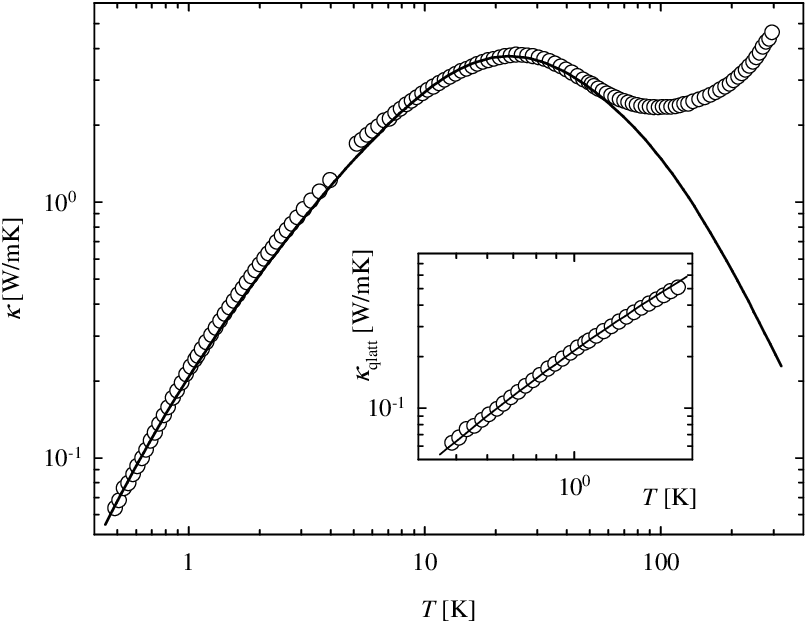}
\caption{\label{fig: Figure 1}
In the main panel, the open circles represent the measured thermal conductivity of \emph{i}-Al$_{72}$Pd$_{19.5}$Mn$_{8.5}$. The solid line shows $\kappa_{Debye}(T)+\kappa_{el}(T)$, where $\kappa_{Debye}(T)$ is obtained by fitting $\kappa_{qlatt}(T)$ to the Debye-type equation (\ref{eq:Debye}) in the temperature range $0.4$ K $<T<40$ K, with $\tau(x)$ from equation (\ref{eq:tau fit}), and $\kappa_{el}(T)=\kappa_{el}^{WF}(T)$. Inset: The open symbols represent $\kappa_{qlatt}(T)$ and the solid line is a fit to equation (\ref{eq:Debye}), with $\tau(x)$ from equation (\ref{eq:tau fit lowT}) for $0.4$ K $<T<1$ K.
}
\end{figure}
\end{center}

\newpage
\begin{center}
\begin{figure}[t]
\includegraphics[width=120mm]{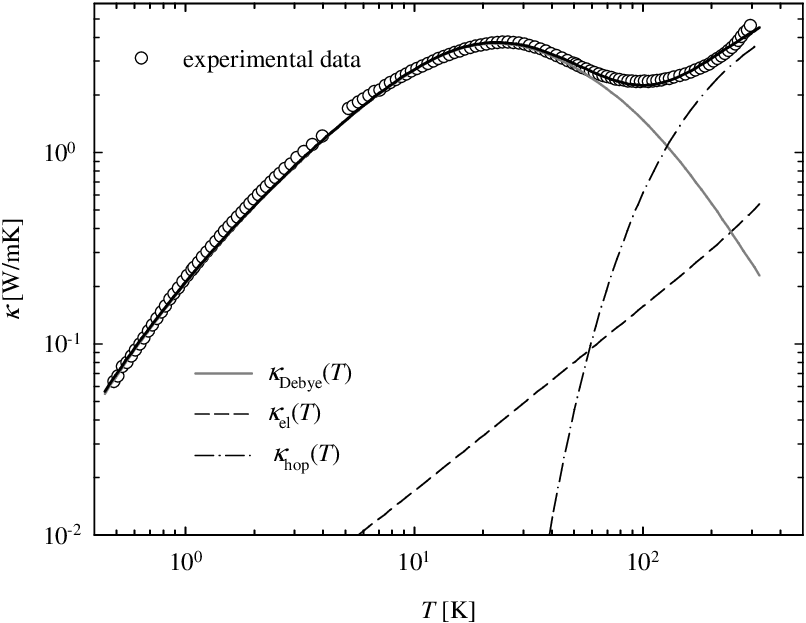}
\caption{\label{fig: Figure 2}
Temperature dependence of the thermal conductivity of \emph{i}-Al$_{72}$Pd$_{19.5}$Mn$_{8.5}$. The black solid line is a fit using equation (\ref{eq:kappa_tot}): the solid grey line represents $\kappa_{Debye}$ from eq. (\ref{eq:Debye}), the dashed line displays $\kappa_{el}(T)$ (see text), and the dotted-dashed line is the hopping contribution given by equation (\ref{eq:activation}).
}
\end{figure}
\end{center}

\newpage
\begin{center}
\begin{figure}[t]
\includegraphics[width=120mm]{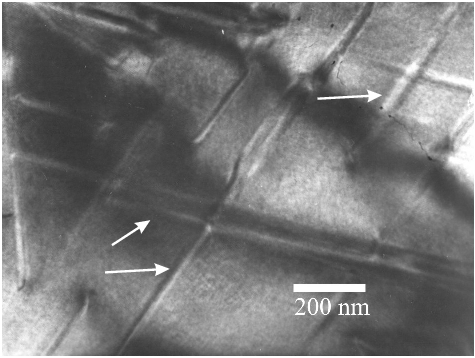}
\caption{\label{fig: Figure 3}
TEM photograph of \emph{i}-Al$_{72}$Pd$_{19.5}$Mn$_{8.5}$. White arrows indicate stacking-fault contrasts.
}
\end{figure}
\end{center}


\begin{thebibliography}{00}

\bibitem{philips72} W.A. Philips, J. Low Temp. Phys. 7 (1972) 351.

\bibitem{anderson72} P.W. Anderson, B.I.  Halperin, C.M. Varma, Philos. Mag.  25 (1972) 1.

\bibitem{chernikov95} M.A. Chernikov, A. Bianchi, H.R. Ott, Phys. Rev. B 51 (1995) 153.

\bibitem{chernikov98} M.A. Chernikov, E. Felder, A.D. Bianchi, C. W\"{a}lti, M. Kenzelmann, H.R. Ott, K. Edagawa, M. de Boissieu, C. Janot, M. Feuerbacher, N. Tamura, K. Urban,  Quasicrystals, Proceedings of the 6th International Conference, Tokyo, Japan 1998, World Scientific (Singapore, 1998) 451.

\bibitem{bert02} F. Bert, G. Bellessa, B. Grushko Phys. Rev. Lett 88 (2002) 255901.

\bibitem{bert01} F. Bert, G. Bellessa, Phys. Rev. B 65 (2001) 014202.

\bibitem{bert00} F. Bert, G. Bellessa, A. Quivy, Y. Calvayrac, Phys. Rev. B 61 (2000) 32.

\bibitem{thompson00} E. Thompson, P.D. Vu, R.O. Pohl, Phys. Rev. B 62 (2000) 11437.

\bibitem{legault98} S. Legault, B. Ellman, J. Str{\"o}m-Olsen, L. Taillefer, T. Lagrasso, D. Delaney, Quasicrystals, Proceedings of the 6th International Conference, Tokyo, Japan 1998, World Scientific (Singapore, 1998) 475.

\bibitem{sterzel00} R. Sterzel, C. Hinkel, A. Haas, A. Langsdorf, G. Bruls, W. Assmus, Europhys. Lett. 49 (2000) 742.

\bibitem{gianno00} K. Giann\`{o}, A.V. Sologubenko, M.A. Chernikov, H.R. Ott, P.C. Canfield, I.R. Fisher, Phys. Rev. B 62 (2000) 292.

\bibitem{fisher99} I.R. Fisher, M.J. Kramers, T.A. Wiener, Z. Islam, A.R. Ross, T.A. Lograsso, A. Kracher, A.I. Goldman, P.C. Canfield, Phil. Mag. B 79 (1999) 1673.

\bibitem{ashcroft-mermin} N.W. Ashcroft, N.D. Mermin Solid State Physics, Saunders College Publishing (1976) 322.

\bibitem{landauro01} C.V. Landauro, H. Solbrig, Physica B 301 (2001) 267.

\bibitem{dolinsek02} J. Dolin\v{s}ek, M. Klanj\v{s}ek, Z. Jagli\v{c}i\'{c}, A. Bilu\v{s}i\'{c}, A. Smontara, J. Phys.: Condens. Matter 14 (2002) 6975.

\bibitem{janot93} C. Janot, A. Magerl, B. Frick, M. de Boissieu, Phys. Rev. Lett. 71 (1993) 871.

\bibitem{boissieu93} M. de Boissieu, M. Boudard, R. Bellisent, M. Quilichini, B. Henion, R. Currat, A.I. Goldman, C. Janot, J. Phys.: Condens. Matter 5 (1993) 4945.

\bibitem{boudard95} M. Boudard, M. de Boissieu, S. Kycia, A.I. Goldman, B. Hennion, R. Bellisent, M. Quilichini, R. Currat, C. Janot, J. Phys.: Condens. Matter 7 (1995) 7299.

\bibitem{krisch02} M. Krisch, R.A. Brand, M.A. Chernikov, H.R. Ott,  Phys. Rev. B 65 (2002) 134201.

\bibitem{berman} R. Berman, Thermal Conduction in Solids (1978, Oxford University Press)

\bibitem{amazit92} Y. Amazit, M. de Boissieu, A. Zarembowitch, Europhys. Lett. 20 (1992) 703.

\bibitem{li01} C. Li, Y. Liu, Phys. Rev. B 63 (2001) 064203.

\bibitem{waelti98} C. W\"{a}lti, E. Felder, M.A. Chernikov, H.R. Ott, M. de Boissieu, C. Janot, Phys. Rev. B 57 (1998) 10504.

\bibitem{klemens} P.G. Klemens, Solid State Physics: Advances in Research and Applications 7 (1958, New York: Academic Press Inc.) 1.

\bibitem{kalugin96} P.A. Kalugin, M.A. Chernikov, A. Bianchi, H.R. Ott, Phys. Rev. B 53 (1996) 14145.

\bibitem{dolinsek05} J. Dolin\v{s}ek, P. Jegli\v{c}, P. McGuiness, Z. Jagli\v{c}i\'{c}, A. Bilu\v{s}i\'{c}, \v{Z}. Bihar, A. Smontara, C. Landauro, M. Feuerbacher, B. Grushko, K.Urban, Phys. Rev. B 72 (2005) 064208.

\bibitem{bihar06} \v{Z}. Bihar, A. Bilu\v{s}i\'{c}, J. Lukatela, A. Smontara, P. Jegli\v{c}, P. McGuiness, J. Dolin\v{s}ek, Z. Jagli\v{c}i\'{c}, J. Janovec, V. Demange, J. M. Dubois, J. All. Comp. 407 (2005) 66.

\bibitem{elser96} V. Elser, Philosoph. Mag. B 73 (1996) 641.

\bibitem{gschneider64} K.A. Gschneider, Solid State Physics: Advances in Research and Applocations 16 (1964, New York: Academic Press) 275.

\bibitem{kajiyima00} K. Kajiyima, K. Edagawa, T. Suzuki, S. Takeuchi, Phil. Mag. Lett. 80 (2000) 49.

\bibitem{swenson04} C.A. Swenson, T.A. Lograsso, N.E. Anderson, Jr., A.R. Ross, Phys. Rev. B 70 (2004) 094201.

\bibitem{pohl02} R.O. Pohl, X. Liu, E. Thompson, Rev. Mod. Phys. 74 (2002) 991.

\end{thebibliography}
\end{document}